\documentstyle[12pt]{article}

\newcommand{\ab}{{(a)}}
\newcommand{\bb}{{(b)}}
\newcommand{\chet}{{(chet)}}
\newcommand{\het}{{(het)}}
\newcommand{\iia}{{(IIA)}}
\newcommand{\be}{\begin{equation}}
\newcommand{\ee}{\end{equation}}
\newcommand{\ben}{\begin{eqnarray}\displaystyle}
\newcommand{\een}{\end{eqnarray}}
\newcommand{\refb}[1]{(\ref{#1})}
\newcommand{\p}{\partial}
\newcommand{\sectiono}[1]{\section{#1}\setcounter{equation}{0}}

\title{STRING STRING DUALITY CONJECTURE IN SIX DIMENSIONS AND CHARGED
SOLITONIC STRINGS}

\author{Ashoke Sen \\
International Centre for Theoretical Physics \\
P.O. Box 586, I-34100 Trieste, ITALY \\
\\
\centerline{and}
\\
Tata Institute of Fundamental Research \\
Homi Bhabha Road, Bombay 400005, INDIA \\
sen@theory.tifr.res.in, sen@tifrvax.bitnet
}

\begin{document}

\maketitle

\vfill

\vbox{\hbox{TIFR-TH-95-16}\hbox{hep-th/9504027}\hbox{April, 1995}}
\hfill ~

\eject

\begin{abstract}

It has recently been conjectured that the type IIA string theory
compactified on K3 and the heterotic string theory compactified on
a four dimensional torus describe identical string theories. The
fundamental heterotic string can be regarded as a non-singular
soliton solution of the type IIA string theory with a semi-infinite
throat. We show that this solution admits 24 parameter
non-singular deformation describing a fundamental heterotic string
carrying electric charge and current. The charge is generated due to the
coupling of the gauge fields to the anti-symmetric tensor field,
and not to an explicit source term. This clarifies
how soliton solutions carrying charge under the Ramond-Ramond fields can
be constructed in the type IIA theory, and provides further support to
the string string duality conjecture. Similarly, the fundamental type
IIA string can be regarded as a non-singular solution of the heterotic
string theory with a semi-infinite throat, but this solution
does not admit any deformation representing charged string. This is
also consistent with the expectation that a fundamental type IIA
string does not carry any charge that couples to the fields originating
in the Ramond-Ramond sector.

\end{abstract}

\eject

\sectiono{Introduction and Summary}

It has recently been conjectured\cite{HULL,DUFF,WITTEN} that the type
IIA string theory compactified on the K3 surface and heterotic string
theory compactified on a four dimensional torus describe identical six
dimensional string theories. Some of the compelling evidences for this
conjecture come from the fact that the two theories give rise to
identical low energy effective field theories, and have identical
moduli spaces\cite{SEIBERG,ASPIN}. If this conjecture is correct, then
the S-duality of the heterotic string theory
compactified on a six dimensional torus\cite{MONTOLIVE}-\cite{GIVEON}
follows as a consequence of the T-duality of the type IIA string
theory\cite{DUFF,WITTEN}. This duality has been called string-string
duality conjecture\cite{DUFF,WITTEN} and we shall refer to it by
this name in this paper. Other aspects of duality symmetries in
string theory relating strong and weak coupling limits have been
discussed in refs.\cite{TOWN,GATES,BARS}.

In order to test this conjecture, we must show that the spectrum of
Bogomol'nyi saturated states\cite{OLIVEWIT}
in the two theories are identical. Here we encounter a puzzle. The
spectrum of the heterotic string theory contains charged Bogomol'nyi
saturated states. On the other hand,
since in the type IIA string theory all the
gauge fields arise from the Ramond-Ramond sector, and the elementary
string states in the theory do not carry any charge under the
gauge fields in the Ramond-Ramond sector, all the elementary string
states in the type IIA string theory are charge neutral. This seems
to contradict the conjecture that the two theories are equivalent.
However, an analysis of the Bogomol'nyi bound formula shows that in the
type IIA string theory, the states that are charged under the gauge field
have masses inversely proportional to the coupling constant of the
type IIA theory\cite{WITTEN}. This shows that we do not expect these
states to arise in the elementary string spectrum, instead they must
arise as solitons in this theory. Extremal charged black holes have
been proposed as possible candidates\cite{HULL}.

The situation however still remains a bit unsatisfactory, since even
for extremal black holes, in order for it to carry electric charge, we
need to put a source of electric field (electric charge) at the
center of the black hole. In other words, if we were not motivated
by the string-string duality conjecture, and were just analyzing the
spectrum of the type IIA string theory compactified on K3, there
would be no reason to include the extremal charged black holes as
solutions in the theory, since they require using charged particles
as sources which are not present in the theory to start with. It
would be much more compelling if we could construct solitons in the
type IIA theory which carry electric charge even if we do not use
an explicit source for the electric field. These solutions will
be analogous to the 't Hooft - Polyakov monopole solutions in
Yang-Mills theories, which give rise to magnetically charged solitons
without the necessity of putting an external source of magnetic field.

One might think that it is impossible to construct a charged soliton
in type IIA string theory without putting a source of electric charge
somewhere. Since all the fields appearing in the low energy effective
action are neutral, the standard Gauss' law will tell us that
the total flux of electric field through a closed surface vanishes
unless there is a source of electric charge somewhere inside the
surface. The reason that we can get around this argument is that the
standard Gauss' law does not apply to the field equations of the
supergravity theory describing the string theory under consideration.
Due to the non-standard coupling of the gauge fields to the rank two
anti-symmetric field in the theory,
the divergence of the electric field is not
zero, but is proportional to the three form field strength contracted
with the electromagnetic field. This allows us to construct charged
solitons in this theory without the necessity of introducing a
source of electric field.

The charged solitons that we shall construct are not particle like
states, but string like states. To this end, we note that the low
energy supergravity theory admits two neutral string like
solutions\cite{DUFF}.
One of them describes the fundamental heterotic string solution of
ref.\cite{DABHOL}, whereas the other describes the fundamental type
IIA string solution. When expressed in the natural variables of the
type IIA string theory, the fundamental heterotic string solution is
non-singular\cite{DUFFREV,DUFFTOW}, and describes an extremal
black string with a
semi-infinite throat, with the type IIA string coupling becoming
strong as we go down the throat. On the other hand,
when expressed in the natural variables of the
heterotic string theory, the fundamental type IIA string solution is
non-singular, and describes an extremal black string with a
semi-infinite throat, with the heterotic string coupling becoming
strong as we go down the throat. We shall show that the first
solution has a 24 parameter deformation describing a fundamental
heterotic string carrying electric charge and electric current.
20 of these parameters describe deformations for which electric
charge per unit length of the string
is equal to the electric current, whereas the rest 4 parameters
describe deformations for which the electric charge per unit
length of the string is equal to the negative of the electric
current. This is consistent with the known properties of the
heterotic string, {\it i.e.} that it carries 20 left moving and
4 right moving currents on the world sheet. But more importantly,
we shall see that this electric charge is not generated by any
source term. In fact, as we go down the semi-infinite throat, the total
electrix flux approaches zero, showing that there is no source
of the electric charge at the far end of the throat. The charge is
generated solely due to the coupling to the anti-symmetric tensor
field as mentioned before. On the other
hand, the second solution, representing the fundamental type IIA string,
does not admit any deformation that
converts it to an electrically charged string. This is again
consistent with our general understanding that the fundamental
type IIA
string does not carry any world sheet current which couples to
the gauge fields originating from the Ramond-Ramond sector of
the theory.

The situation may be summarized as follows:

\begin{itemize}

\item {The type IIA string theory contains a non-singular soliton solution
carrying the quantum numbers of a fundamental heterotic string. This
solution admits non-singular 24 parameter deformations which has the
quantum numbers of a fundamental heterotic string carrying 20 left moving
currents and 4 right moving currents on the world sheet. This is consistent
with the known properties of the fundamental heterotic string.}

\item {The heterotic string theory contains a non-singular soliton solution
carrying the quantum numbers of a fundamental type IIA string. This solution
does not admit any deformation which might represent fundamental type IIA
string carrying world sheet currents that couple to the gauge fields. This is
consistent with the known properties of the fundamental type IIA string.}

\end{itemize}

These results provide further support to the string-string duality
conjecture in six dimensions.

We now give the plan of the paper. In section 2 we shall write
down the action of the six dimensional supergravity theory in
two different sets of variables, $-$ one natural to the heterotic
string and the other natural to the type IIA string. We shall
also give the map between these two sets of variables. Finally
we shall write down the solutions describing the fundamental
heterotic string and the fundamental type IIA string in both
sets of variables and discuss the geometrical properties of these
solutions. In section 3 we
shall use the by now standard method of $O(d,d)$
transformation\cite{VENEZ,HASSAN,MACST}
to generate the deformed heterotic string solution describing
the charged string. We show that the deformed solutions are
non-singular, and do not have any source of electric charge at
the far end of the throat. We also show that the $O(d,d)$
transformations {\it cannot} be used to generate a charged
type IIA string solution. We end in section 4 with a few remarks
about the possibility of constructing solvable conformal field
theories that might describe the fundamental heterotic string as a
solution of the type II string theory, and vice versa.

We end this section by stating some of the notations that we shall be
using. The unprimed fields will denote the fields that couple naturally
to the heterotic string, whereas the primed fields will denote the
fields which couple naturally to the type IIA string. The subscript
$\het$ will be used to denote the solution that represents a
fundamental heterotic string, whereas the subscript $\iia$ will denote
the solution representing a fundamental type IIA string. Finally,
the subscript $\chet$ will denote the solution representing a heterotic
string carrying electric charge and current.

\sectiono{Low Energy Effective Action, and Solitonic Strings}

The material in this section will be a review of the results contained
in refs.\cite{SEIBERG,HULL,DUFF,WITTEN}. The low energy effective action
describing the heterotic string compactified on a four dimensional torus
($T^4$) is given by,
\ben \label{e1}
S &\propto& \int d^6 x \sqrt{-G} e^{-\Phi} [R_G + G^{\mu\nu} \p_\mu \Phi
\p_\nu\Phi -{1\over 12} G^{\mu\mu'} G^{\nu\nu'} G^{\rho\rho'}
H_{\mu\nu\rho} H_{\mu'\nu'\rho'} \nonumber \\
&& - G^{\mu\mu'} G^{\nu\nu'} F^\ab_{\mu\nu} (LML)_{ab} F^\bb_{\mu'\nu'}
+{1\over 8} G^{\mu\nu} Tr(\p_\mu M L \p_\nu M L)] \, ,
\een
where $G_{\mu\nu}$, $B_{\mu\nu}$, $A_\mu^\ab$ ($0\le \mu \le 5$,
$1\le a\le 24$), $\Phi$ and $M$  denote respectively the
string metric, the antisymmetric tensor field, the 24 abelian gauge fields,
the dilaton field, and the $24\times 24$ matrix valued scalar field
representing an element of $O(4,20)/(O(4)\times O(20))$, satisfying
\be \label{e3}
M^T=M, \qquad MLM^T = L\, ,
\ee
where,
\be \label{e4}
L = \pmatrix{ -I_{20} & \cr & I_4 \cr} \, .
\ee
$R_G$ denotes the scalar curvature associated with the metric $G_{\mu\nu}$,
and,
\ben \label{e2}
F^\ab_{\mu\nu} & = & \p_\mu A^\ab_\nu - \p_\nu A^\ab_\mu \nonumber \\
H_{\mu\nu\rho} & = & (\p_\mu B_{\nu\rho} + 2 A^\ab_\mu L_{ab} F^\bb_{\nu
\rho}) +\hbox{ cyclic permutations of $\mu,\nu,\rho$} \, .
\een
The matrix valued scalar field $M$ originate from the internal components
of the ten dimensional metric, antisymmetric tensor field and the gauge
fields. Four of the 24 gauge fields originate from the components of the
ten dimensional metric, four of them originate from the components of the
ten dimensional anti-symmetric tensor field, and the remaining sixteen
gauge fields can be identified to the gauge fields associated with the
Cartan subalgebra of $E_8\times E_8$ or $SO(32)$.

On the other hand, the low energy effective action describing the type IIA
string compactified on $K3$ is given by,
\ben \label{e5}
S' &\propto& \int d^6 x \Big( \sqrt{-G'} [e^{-\Phi'} \{ R_G' + G^{\prime
\mu\nu} \p_\mu \Phi' \p_\nu\Phi'
-{1\over 12} G^{\prime\mu\mu'} G^{\prime\nu\nu'} G^{
\prime\rho\rho'}
H'_{\mu\nu\rho} H'_{\mu'\nu'\rho'} \nonumber \\
&& +{1\over 8} G^{\prime\mu\nu} Tr(\p_\mu M' L \p_\nu M' L)\}
- G^{\prime\mu\mu'} G^{\prime\nu\nu'} F^{\prime \ab}_{\mu\nu}
(LM'L)_{ab} F^{\prime \bb}_{\mu'\nu'} ] \nonumber \\
&& -{1\over 4} \varepsilon^{\mu\nu\rho\sigma\tau\epsilon}
B'_{\mu\nu} F^{\prime \ab}_{\rho\sigma} L_{ab}
F^{\prime \bb}_{\tau\epsilon} \Big) \, ,
\een
where $G'_{\mu\nu}$, $B'_{\mu\nu}$, $A_\mu^{\prime\ab}$,
$\Phi'$ and $M'$  denote respectively the
string metric, the antisymmetric tensor field, the 24 abelian gauge fields,
the dilaton field, and the $24\times 24$ matrix valued scalar field
representing an element of $O(4,20)/(O(4)\times O(20))$. $M'$ satisfies
equations identical to those satisfied by $M$:
\be \label{en1}
M' L M^{\prime T}= L, \qquad M^{\prime T} = M'\, .
\ee
Here
\ben \label{e6}
F^{\prime \ab}_{\mu\nu} & = & \p_\mu A^{\prime \ab}_\nu -
\p_\nu A^{\prime \ab}_\mu \nonumber \\
H'_{\mu\nu\rho} & = & \p_\mu B'_{\nu\rho}
+\hbox{ cyclic permutations of $\mu,\nu,\rho$} \, .
\een
Note that there is no Chern-Simons term involving the gauge fields in
the expression for $H'$. In this case the scalar field degrees of freedom
contained in $M'$ come from the internal components (along the tangent
space of $K3$) of the metric and the anti-symmetric tensor field. On the
other hand, the gauge fields come from various components of the
anti-symmetric tensor field $C_{MNP}$ and the vector field $E_M$
($0\le M\le 9$) in the
ten dimensional theory which originate in the Ramond-Ramond sector of
the type IIA superstring theory. In particular $E_\mu$ gives one gauge
field, the components $C_{mn\mu}$, where $m,n$ denote the tangent space
directions on $K3$, give 22 gauge fields, and dualization of the three
form $C_{\mu\nu\rho}$ in six dimensions give another gauge field.

It can be easily verified that the equations of motion derived from the
actions \refb{e1} and \refb{e5} are identical, provided one makes the
following identification of fields:
\ben \label{e7}
&& \Phi' = -\Phi, \qquad G'_{\mu\nu} = e^{-\Phi} G_{\mu\nu},
\qquad M'=M, \qquad A^{\prime \ab}_\mu = A^\ab_\mu, \nonumber \\
&& \sqrt{-G} e^{-\Phi} H^{\mu\nu\rho} = {1\over 6}
\varepsilon^{\mu\nu\rho\sigma\tau\epsilon} H'_{\sigma\tau\epsilon}\, .
\een

Let us now describe solutions of the equations of motion derived from the
above effective actions, describing fundamental heterotic string, and
fundamental type IIA string respectively. To this end, we note that a
fundamental heterotic string will carry electric type $B_{\mu\nu}$
charge, and hence magnetic type $B'_{\mu\nu}$ charge, since $B_{\mu\nu}$
and $B'_{\mu\nu}$ are related by duality transformations. On the other hand,
the fundamental type IIA string will carry electric type $B'_{\mu\nu}$
charge and magnetic type $B_{\mu\nu}$ charge. Thus they must be represented
by different solutions of the field equations. Both of these solutions are
given in refs.\cite{DUFF,DUFFREV}, and are related to the solitonic
string solutions of ref.\cite{DABHOL} in appropriate variables.
The fundamental heterotic string
solution is given in the unprimed field variables by
\ben \label{e8}
ds^2_\het &=& \Big(1 + {C\over r^2}\Big)^{-1} (-dt^2 + (dx^5)^2)
+ dr^2 + r^2 d\Omega_3^2 \, , \nonumber \\
e^{-\Phi_\het} &=& 1 + {C\over r^2}\, , \nonumber \\
B_{\het 5t} &=& {C\over C + r^2}\, , \nonumber \\
A^\ab_{\het\mu} &=& 0, \qquad M_\het = I_{24} \, ,
\een
where $d\Omega_3$ denotes the line element on a three sphere $S^3$, and
$C$ is a constant determined by the heterotic string tension.
This solution is identical to the one constructed by
Dabholkar et. al.\cite{DABHOL}.
The metric has a singularity at $r=0$ in these variables. However, the same
solution expressed in terms of the primed variables is given by,
\ben \label{e9}
ds^{\prime 2}_\het &=& -dt^2 + (dx^5)^2
+ \Big( 1 + {C \over r^2} \Big) dr^2 + (C + r^2) d\Omega_3^2 \, , \nonumber \\
e^{-\Phi'_\het} &=& \Big( 1 + {C\over r^2}\Big)^{-1} \, , \nonumber \\
H'_{\het ijk} &=& - 2 C \varepsilon_{ijk} \, , \nonumber \\
A^{\prime \ab}_{\het\mu} &=& 0, \qquad M'_\het = I_{24} \, ,
\een
where $\varepsilon_{ijk}$ denotes the volume form on the three sphere
$S^3$.
The apparent singularity of the metric at $r=0$ is only a coordinate
singularity\cite{DUFFREV}, as can be seen by defining new coordinate $\rho$
near $r=0$:
\be \label{e9a}
\rho = \ln r \, .
\ee
In this coordinate, the point $r=0$ is mapped to the point $\rho=-\infty$,
and the metric near $r=0$ takes the form:
\be \label{e10}
ds^{\prime 2}_\het \simeq - dt^2 + (dx^5)^2 + C d\rho^2 + C d\Omega_3^2
\ee
which represents the geometry of a semi-infinite line labelled by $\rho$
tensored with a three sphere of constant radius, and a two dimensional
Minkowski space labelled by $t$ and $x^5$. The dilaton near $r\simeq 0$
takes the form:
\be \label{e10a}
e^{-\Phi'_\het} \simeq C e^{2\rho}
\ee
showing that the string coupling grows as we go down the semi-infinite
throat labelled by $\rho$ towards $\rho=-\infty$.
This shows that the fundamental heterotic string can be represented as a
non-singular soliton solution of the type IIA string theory compactified
on $K3$.

Similarly, we can construct a solution of the equations of motion of the
six dimensional effective field theory which represents the fundamental
type IIA string. In the primed variables this solution is again identical
to the one constructed in ref.\cite{DABHOL}, and is given by,
\ben \label{e11}
ds^{\prime 2}_\iia &=& \Big(1 + {C'\over r^2}\Big)^{-1} (-dt^2 + (dx^5)^2)
+ dr^2 + r^2 d\Omega_3^2 \, , \nonumber \\
e^{-\Phi'_\iia} &=& 1 + {C'\over r^2}\, , \nonumber \\
B'_{\iia 5t} &=& {C'\over C' + r^2}\, , \nonumber \\
A^{\prime \ab}_{\iia \mu} &=& 0, \qquad M'_\iia = I_{24} \, ,
\een
where the constant $C'$ is now determined by the string tension of the
type IIA theory. The metric is again singular at the origin. But in terms
of the unprimed variables, the solution looks like,
\ben \label{e12}
ds^{2}_\iia &=& -dt^2 + (dx^5)^2
+ \Big( 1 + {C' \over r^2} \Big) dr^2 + (C' + r^2) d\Omega_3^2 \, ,
\nonumber \\
e^{-\Phi_\iia} &=& \Big( 1 + {C'\over r^2}\Big)^{-1} \, , \nonumber \\
H_{\iia ijk} &=& - 2 C' \varepsilon_{ijk} \, , \nonumber \\
A^{\ab}_{\iia \mu} &=& 0, \qquad M_\iia = I_{24} \, ,
\een
which can again be seen to be non-singular. Thus we see that
the type IIA string can be regarded as a non-singular soliton solution of the
heterotic string theory.

\sectiono{Charged Solitonic Strings}

In the last section we saw that the type IIA string theory compactified on
$K3$ contains a non-singular soliton solution which has the quantum numbers
of the heterotic string. We shall now study in more detail whether this
solution has the right properties expected of a heterotic string. In
particular, we know that the heterotic string can carry 20 left moving
and 4 right moving world-sheet currents that couple to the 24 gauge fields.
So in order to interprete this soliton solution as the fundamental heterotic
string, we must show that the solution admits a 24 parameter deformation
which represents a heterotic string carrying 20 left moving and 4 right
moving currents on the world sheet. (Since we are considering a static string
extending in the $x^5$ direction, in this case the world sheet coordinates
can be identified to the space-time coordinates $x^5$ and $t$.)

But here we encounter a puzzle. It is well known that the type II string
does not have any elementary excitations that carry electric charge
associated with the Ramond-Ramond fields. Since all the gauge fields
in the compactified theory arise from the Ramond-Ramond sector, this
means that the theory does not have any state (field) that carries
gauge charges. Thus it would seem impossible to construct a soliton
solution
in the theory that carries gauge charges. On the other hand, this is
precisely what we need to do if we have to construct solitonic strings
representing heterotic strings.

A resolution to this puzzle comes from looking at the equations of motion
for the gauge fields. In terms of the primed variables, these equations
take the form:
\be \label{e30}
D'_\nu [ (M' L)_{ab} F^{\prime \bb \mu \nu}] + {1\over 12} (\sqrt{-G})^{-1}
\varepsilon^{\tau \epsilon \mu \sigma\rho\nu} H'_{\sigma\rho\nu}
F^{\prime \ab}_{\tau\epsilon} = 0
\ee
This shows that although the theory does not contain any charged field,
the second term in the above equation can act as a source term in the
gauge field equations of motion, and give rise to solitons carrying net
electric charge. As we shall see this is precisely the mechanism that
allows us to construct a charged heterotic string solution.

The explicit construction of the solution is done using the solution
generating transformations\cite{VENEZ}-\cite{TORBH}. The
relevant group of transformations belong
to the coset $(O(20,1)/O(20))\times (O(4,1)/O(4))$.
Since the transformations used here are identical
to the ones used in ref.\cite{TORBH}, we shall not give the details here,
but only quote the final result. The final solution is characterized by
24 extra parameters, consisting of two boost angles $\alpha$ and $\beta$,
a 20 dimensional unit vector $\vec n$, and a 4 dimensional unit vector
$\vec p$. In the unprimed variables (where the solution generating
transformations act naturally) the transformed solution is given by,
\ben \label{e13}
ds^2_\chet &=& r^2 \Delta^{-1} [ -(r^2+C) dt^2 + C (\cosh\alpha -
\cosh\beta) dt dx^5  \nonumber \\
&& + (r^2 + C \cosh\alpha \cosh\beta) (dx^5)^2]
\nonumber \\
&& + (dr^2 + r^2 d\Omega_3^2) \, ,
\een
\be \label{e14}
B_{\chet 5t} = {C\over 2\Delta} (\cosh\alpha + \cosh\beta)
\{ r^2 + {1\over 2} C (1 + \cosh\alpha \cosh \beta) \} \, ,
\ee
\be \label {e15}
e^{-\Phi_\chet} = {\Delta^{1/2} \over r^2} \, ,
\ee
\ben \label{e16}
A^\ab_{\chet t} &=& -{n^\ab \over 2 \sqrt 2 \Delta} C \sinh \alpha
\{ r^2 \cosh\beta + {1\over 2} C (\cosh\alpha + \cosh \beta) \}
\nonumber \\
&& \qquad \qquad \hbox{for} \qquad 1 \le a \le 20 \, , \nonumber \\
&=& -{p^{(a-20)} \over 2 \sqrt 2 \Delta} C \sinh \beta
\{ r^2 \cosh\alpha + {1\over 2} C (\cosh\alpha + \cosh \beta) \}
\nonumber \\
&& \qquad \qquad \hbox{for} \qquad 21 \le a \le 24 \, , \nonumber \\
\een
\ben \label{e17}
A^\ab_{\chet 5} &=& - {n^\ab \over 2\sqrt 2 \Delta} C \sinh \alpha \{ r^2
+ {1\over 2} C \cosh \beta (\cosh\alpha + \cosh \beta) \} \nonumber \\
&& \qquad \qquad \hbox{for} \qquad
1\le a \le 20 \, , \nonumber \\
&=&  {p^{(a-20)} \over 2\sqrt 2 \Delta} C \sinh \beta \{ r^2
+ {1\over 2} C \cosh \alpha (\cosh\alpha + \cosh \beta) \} \nonumber \\
&& \qquad  \qquad \hbox{for} \qquad
21\le a \le 24 \, , \nonumber \\
\een
\be \label{e18}
M = I_{28} + \pmatrix{ P nn^T & Q np^T \cr Q pn^T & P pp^T \cr} \, ,
\ee
where,
\be \label{e19}
\Delta = r^4 + Cr^2 ( 1 + \cosh\alpha \cosh\beta) + {C^2 \over 4}
(\cosh\alpha + \cosh\beta)^2 \, ,
\ee
\be \label{e19a}
P = {C^2 \over 2\Delta} \sinh^2 \alpha \sinh^2 \beta \, ,
\ee
\be \label{e19b}
Q = - C \Delta^{-1} \sinh\alpha \sinh\beta \{ r^2 + {1\over 2} C
(1 + \cosh\alpha \cosh \beta) \} \, ,
\ee
For $\beta=0$, these solutions agree with the charged string solutions
found in ref.\cite{MACST}.

By examining the solution we see that the string tension of the
heterotic string is now proportional to $C(\cosh\alpha + \cosh\beta)$.
Also the solution is as usual singular at $r=0$.
We shall now verify that the solution is non-singular in the primed
variables. In these variables, the solution takes the form:
\ben \label{e24}
ds^{\prime 2}_\chet &=& \Delta^{-1/2} [ -(r^2+C) dt^2 + C (\cosh\alpha -
\cosh\beta) dt dx^5 \nonumber \\
&& + (r^2 + C \cosh\alpha \cosh\beta) (dx^5)^2]
\nonumber \\
&& + \Delta^{1/2} \Big( {dr^2 \over r^2} + d\Omega_3^2 \Big)  \, ,
\een
\be \label{e25}
e^{-\Phi'_\chet} = {r^2 \over \Delta^{1/2}} \, ,
\ee
\be \label{e26}
H'_{\chet ijk} = - C (\cosh\alpha + \cosh \beta) \varepsilon_{ijk} \, ,
\ee
\be \label{e27}
M'_\chet = M_\chet \, ,
\ee
\be \label{e28}
A^{\prime \ab}_{\chet \mu} = A^\ab_{\chet\mu} \, .
\ee
Near $r=0$, we again use the coordinate $\rho=\ln r$. In this coordinate
system the metric near $r=0$ takes the form:
\ben \label{e29}
ds^{\prime 2}_\chet &\simeq& {2 \over \cosh\alpha + \cosh \beta} \Big(
- dt^2 + \cosh\alpha \cosh \beta (dx^5)^2 \nonumber \\
&& + (\cosh \alpha - \cosh \beta) dt dx^5 \Big)
+ {C \over 2} (\cosh\alpha + \cosh \beta) (d\rho^2 + d \Omega_3^2)\, ,
\nonumber \\
\een
and describes a completely non-singular geometry. We also see from
eqs.\refb{e16}, \refb{e17} that the
gauge fields are non-singular as $r\to 0$.

We define the electric charge per unit length ($q^\ab$)
and electric current ($j^\ab$) carried
by the solution in terms of the asymptotic values of the gauge fields
$(r\to \infty)$:
\be \label{e20}
F^{\prime \ab}_{rt} \simeq {q^\ab \over r^3}, \qquad F^{\prime \ab}_{r5}
\simeq {j^\ab \over r^3}\, .
\ee
Comparing eq.\refb{e20} with eqs.\refb{e16}, \refb{e17} and \refb{e28}
we get
\ben \label{e21}
q^\ab &=& {n^\ab \over \sqrt 2} C \sinh \alpha \cosh \beta \, , \qquad
\hbox{for} \qquad 1\le a \le 20, \nonumber \\
&=& {p^{(a-20)} \over \sqrt 2} C \sinh \beta \cosh \alpha \, , \qquad
\hbox{for} \qquad 21\le a \le 24,
\een
and,
\ben \label{e22}
j^\ab &=& {n^\ab \over \sqrt 2} C \sinh \alpha \, , \qquad
\hbox{for} \qquad 1\le a \le 20, \nonumber \\
&=& - {p^{(a-20)} \over \sqrt 2} C \sinh \beta \, , \qquad
\hbox{for} \qquad 21\le a \le 24\, .
\een
Thus, for small $\alpha$, $\beta$,
\ben \label{e23}
q^\ab = j^\ab & \simeq & {n^\ab \over \sqrt 2} C \alpha \qquad \hbox{for}
\qquad 1 \le a \le 20 \, , \nonumber \\
q^\ab = - j^\ab & \simeq & {p^{(a-20)} \over \sqrt 2} C \beta
\qquad \hbox{for}
\qquad 21 \le a \le 24 \, .
\een
This shows that the deformed solution does represent a fundamental heterotic
string carrying 20 left-moving and 4 right moving currents on its
world-sheet.

Let us briefly examine the source of the electric charge. From eq.\refb{e16},
\refb{e17} we can easily verify that the total electric flux per unit length
of the string through a surface of constant $r$ (or $\rho$) vanishes as
$r\to 0$ ($\rho \to -\infty$).
This shows that there is no source of electric field hidden at the far end
of the semi-infinite geometry ($\rho\to -\infty$). In fact, a more detailed
examination of the solution shows that the total electric charge carried
by the solution is indeed given by the integral of the second term in
eq. \refb{e30} over the whole space, therby proving that this is the only
source responsible for the electric charge of the soliton.

Let us now ask the opposite question: do we get charged type IIA string
soliton by starting with the neutral type IIA string soliton, and applying
the solution generating transformations on this solution? If the answer is
yes, it would imply that the type IIA string soliton {\it does not} satisfy
the properties of a fundamental type IIA string, since it is known that
the fundamental type IIA string does not carry any world-sheet current that
couples to the gauge fields originating from the Ramond-Ramond sector.
We again start from the type IIA soliton expressed in the unprimed variables,
since the solution generating transformations act naturally on the
unprimed variables. This solution has been given in eq.\refb{e12}. Notice
that the the part of the metric involving $(-dt^2 + (dx^5)^2)$ does not
have any conformal factor. This, in turn, implies that this solution is
left invariant under the solution generating transformations, and we do not
generate any new solutions! This is precisely what we want, since if we had
gotten a new solution using the solution generating transformation, it would
have almost certainly corresponded to a charge carrying solution, thereby
showing that
the soliton does not have the right properties for being identified as the
fundamental type IIA string. When expressed in the primed variables, the
$(-dt^2 + (dx^5)^2)$ term in the metric does have a conformal factor, but
no solution generating transformation is known that acts naturally on the
primed variables, and most likely no such transformation exists. Thus we
conclude that the type IIA string soliton, which is a non-singular solution
of the heterotic string theory, is rigid, in that it does not allow
deformations which correspond to charge carrying strings.

\sectiono{String Solitons as Exact Conformal Field Theories}

We have seen in the previous two sections that the fundamental heterotic
string can be regarded as a non-singular solution of the type IIA string
theory and vice versa. It is natural to ask if it is possible to represent
them as exact conformal field theories. Unfortunately the complete answer
to this question is not known. However by examining the (uncharged) solutions
near the semi-infinite geometry ($r\simeq 0$) we see that it can be
represented as SO(3) WZW model with a linear dilaton background\cite{CALLAN}.
The level of the WZW theory will depend on the precise relationship
between the coupling constants of the two string theories.
In particular, the product of the coupling constants of the two theories
satisfy a quantization condition\cite{DUFF,DUFFREV}, and in the case where
the quantized product takes its minimal value we would expect both the
solutions to be described by a level 1 SO(3) WZW model.


\begin{thebibliography}{99}

\bibitem{HULL}
C. Hull and P. Townsend, preprint QMW-94-30 (hep-th/9410167).

\bibitem{DUFF}
M. Duff, preprint CTP-TAMU-49/94 (hep-th/9501030); \\
M. Duff and R. Khuri, Nucl. Phys. {\bf B411} (1994) 473 (hep-th/9305142).

\bibitem{WITTEN}
E. Witten, preprint IASSNS-HEP-95-18 (hep-th/9503124).

\bibitem{SEIBERG}
N. Seiberg, Nucl. Phys. {\bf B303} (1988) 286.

\bibitem{ASPIN}
P. Aspinwall and D. Morrison, preprint DUK-TH-94-68 (hep-th/9404151).


\bibitem{MONTOLIVE}
C. Montonen and D. Olive, Phys. Lett. {\bf B72} (1977) 117; \\
P. Goddard, J. Nyuts and D. Olive, Nucl. Phys. {\bf B125} (1977) 1; \\
H. Osborn, Phys. Lett. {\bf B83} (1979) 321.

\bibitem{FONT}
A. Font, L. Ibanez, D. Lust and F. Quevedo, Phys. Lett. {\bf B249}
(1990) 35; \\
S.J. Rey, Phys. Rev. {\bf D43} (1991) 526; \\
S. Kalara and D. Nanopoulos, Phys. Lett. {\bf B267} (1991) 343.

\bibitem{SREV}
A. Sen, Int. J. Mod. Phys. {\bf A9} (1994) 3707 (hep-th/9402002),
and references therein.

\bibitem{JHSREV}
J. Schwarz, preprint CALT-68-1965 (hep-th/9411178).

\bibitem{BOUND}
A. Sen, Phys. Lett. {\bf B329} (1994) 217 (hep-th/9402032).

\bibitem{GAUNTHAR}
J. Gauntlett and J. Harvey, preprint EFI-94-36 (hep-th/9407111).

\bibitem{VAFAWITTEN}
C. Vafa and E. Witten, preprint HUTP-94-A017 (hep-th/9408074).

\bibitem{SEGAL}
G. Segal, to appear.

\bibitem{GIVEON}
L. Girardello, A. Giveon, M. Porrati and A. Zaffaroni,
preprint NYU-TH-94/06/02 (hep-th/9406128);
preprint NYU-TH-94/12/01 (hep-th/9502057).

\bibitem{TOWN}
P. Townsend, preprint DAMTP-R/95/2 (hep-th/9501068).

\bibitem{GATES}
S. Gates and V. Rodgers, UMD EPP 95-76 (hep-th/9503237).

\bibitem{BARS}
I. Bars, preprint USC-95/HEP-82 (hep-th/9503228).

\bibitem{OLIVEWIT}
E. Witten and D. Olive, Phys. Lett. {\bf B78} (1978) 97.

\bibitem{DABHOL}
A. Dabholkar, G. Gibbons, J. Harvey and F. Ruiz, Nucl. Phys. {\bf B340}
(1990) 33.

\bibitem{DUFFREV}
M. Duff, R. Khuri and J. Lu, preprint CTP/TAMU-67/92 (hep-th/9412184),
and references therein.

\bibitem{DUFFTOW}
M. Duff, G. Gibbons and P. Townsend, preprint DAMTP/R-93/5
(hep-th/9405124).

\bibitem{VENEZ}
G. Veneziano, Phys. Lett. {\bf B265} (1991) 287; \\
K. Meissner and G. Veneziano, Phys. Lett. {\bf B267} (1991) 33;
Mod. Phys. Lett. {\bf A6} (1991) 3397; \\
A. Sen, Phys. Lett. {\bf B271} (1991) 295; {\bf B274} (1991) 34; \\
M. Gasperini, J. Maharana and G. Veneziano, Phys. Lett. {\bf B272}
(1991) 272.

\bibitem{HASSAN}
S. Hassan and A. Sen, Nucl. Phys. {\bf B375} (1992) 103
(hep-th/9109038); \\
S. Hassan, preprint TIFR-TH-94-26 (hep-th/9408060).

\bibitem{MACST}
A. Sen, Nucl. Phys. {\bf B388} (1992) 457.

\bibitem{TORBH}
A. Sen, preprint TIFR-TH-94-47  (hep-th/9411187).

\bibitem{CALLAN}
C. Callan, J. Harvey and A. Strominger, Nucl. Phys. {\bf B359} (1991)
611; {\bf B367} (1991) 60.


\end{thebibliography}
\end{document}